\documentclass[%
aip,
amsmath,amssymb,
reprint,%
]{revtex4-1}

\usepackage{graphicx}
\usepackage{dcolumn}
\usepackage{bm}

\usepackage[utf8]{inputenc}
\usepackage[T1]{fontenc}
\usepackage{mathptmx}
\usepackage{etoolbox}
\usepackage{color}
\usepackage{autobreak}

\makeatletter
\def\@email#1#2{%
	\endgroup
	\patchcmd{\titleblock@produce}
	{\frontmatter@RRAPformat}
	{\frontmatter@RRAPformat{\produce@RRAP{*#1\href{mailto:#2}{#2}}}\frontmatter@RRAPformat}
	{}{}
}%
\makeatother
\begin{document}
	
	\preprint{AIP/123-QED}
	
	\title[]{Exceptional Point Modulated by Kerr Effect in Anti-Parity-Time Symmetry System}
	
	\affiliation{ 
		School of Physics, Beihang University, Beijing 100191, China
	}%
	\affiliation{%
		Research Institute of Frontier Science, Beihang University, Beijing, 100191, China
	}%
	\affiliation{ 
		Beijing Academy of Quantum Information Sciences, Beijing 100193, China
	}%

	\author{Tuoyu Chen}
	\affiliation{ 
		School of Physics, Beihang University, Beijing 100191, China
	}%
	\author{Zhisong Xiao}%
	\affiliation{ 
		School of Physics, Beihang University, Beijing 100191, China
	}%
	\affiliation{%
		Research Institute of Frontier Science, Beihang University, Beijing, 100191, China
	}%
	\affiliation{ 
		Beijing Academy of Quantum Information Sciences, Beijing 100193, China
	}%
	\author{Shuo jiang}%
	\author{Wenxiu Li}%
	\author{Jincheng Li}%
	\author{Yuefei Wang}%
	\author{Xiaochen Wang}%
	\author{Anping Huang}
	\affiliation{ 
		School of Physics, Beihang University, Beijing 100191, China
	}%
	
	\author{Hao Zhang$^*,$}
	\email{haozhang@buaa.edu.cn.}
	\affiliation{%
		Research Institute of Frontier Science, Beihang University, Beijing, 100191, China
	}%

	
\begin{abstract}
With respect to parity-time (PT) symmetry, anti-parity-time (APT) symmetric system exhibits much easier readout mechanism due to its real frequency splitting. Generally, such
systems need to be operated at exceptional points (EPs) to obtain the best performance. However, strict conditons to locate APT symmetric systems at their EPs precisely put restraints on their practical applications. To overcome this problem, we propose a scheme to manipulate the EPs in APT symmetric configuration by Kerr effect. It is demonstrated that operating EPs by self-phase modulation alone will impede the frequency splitting caused by external perturbations, while cross-phase modulation can enhance the response to measurable perturbations. We also investigate the thermal effect induced by high light intensity, which could reduce the power to manipulate EPs. This proposed scheme can pave a new way in fabricating devices based on APT symmetry.
\end{abstract}
	
\maketitle
	
Non-Hermitian optical systems operating near exceptional points (EPs) have received considerable attention due to nontrivial phenomena related to them , such as loss-induced transparency\cite{guo2009observation,peng2014loss}, nonreciprocal light transmission\cite{feng2011nonreciprocal,sukhorukov2010nonlinear,ramezani2010unidirectional,yin2013unidirectional,fan2012comment} and ultrasensitive sensing\cite{chen2017exceptional,wiersig2014enhancing,wiersig2016sensors,hodaei2017enhanced,sunada2017large}. EPs are non-Hermitian degeneracies at which the eigenvalues and their corresponding eigenstates coalesce\cite{chang2014parity,jin2018incident,jin2018parity,wang2020electromagnetically,guo2009observation,ruter2010observation}. They have been implemented in various non-Hermitian optical systems via optical chirality behaviors\cite{chen2017exceptional,wiersig2011structure}, Parity-Time (PT) symmetry\cite{chang2014parity,ruter2010observation,hodaei2014parity,peng2014parity,schnabel2017pt} and anti-Parity-Time (APT) symmetry\cite{yang2017anti,zhang2019dynamically}. Optical chirality arises from unbalanced contribution of clockwise and counter-clockwise travelling wave modes, and PT symmetry demands balance of gain and loss. As a counterpart of PT symmetry, APT symmetry requires a dissipative coupling between two resonant modes with frequency detuning, which does not need gain and can be realized in dissipative systems\cite{yang2017anti}. In addition, APT symmetry  can exhibit a completely real frequency splitting which can be directly measured at
the output power spectrum \cite{li2020real,de2019high}.
	
Despite the tremendous advantages of APT symmetry, devices based on APT symmetry suffer from fabrication errors and optical nonlinearities. In whisper-gallery-mode resonators (WGMRs) with high Q factor and small mode volume, high intensity can be easily induced and the induced high intensity allows for significant Kerr effect due to its ubiquitousness in centrosymmetric materials \cite{lin2017nonlinear,shen2016compensation,brasch2016photonic}. The Kerr nonliearity can have a major impact on the location of EP in APT symmetry, which is determined by the frequency detuning of resonant modes. As a result, locating APT symmetric system at their EPs precisely becomes more difficult and calls for a more stringent control of parameters.
	
Although optical nonlinearity often violates the strict parameter conditions in EP-based devices, it can be utilized to overcome these challenges, such as saturable gain and loss\cite{hassan2015nonlinear} and Kerr effect\cite{ramezani2010unidirectional,ramezanpour2021tuning} in PT symmetry. Along this way, EPs in APT symmetry can be manipulated by Kerr nonlinearity.
	
In this paper, we find that Kerr effect provides additional degree of freedom to control the phase transition in the system. Moreover, a scheme to modulate EPs by taking advantage of Kerr shift is put forward. We demonstrate theoretically that Kerr effect stimulated in the form of self-phase modulation (SPM) impedes frequency splitting while cross-phase modulation (XPM) could be utilized to enhance the frequency splitting. This scheme can be exploited to combine steerable EPs with enhanced response to detectable perturbations.
	
This paper is organized as follows. In SEC.\ref{SEC:2}, a general configuration of APT system is introduced and the linear condition is reviewed in SEC.\ref{SEC:3}. In SEC.\ref{SEC:4} and SEC.\ref{SEC:5}, how Kerr effect influence EPs and their response to measurable perturbations is analyzed, and the influence of thermal effect in the system is given in  SEC.\ref{SEC:6}. This paper is concluded in SEC.\ref{SEC:7}.
\section{Theoretical Model} \label{SEC:2}
\begin{figure}[htbp]
	\centering
	\includegraphics[width=8.5cm]{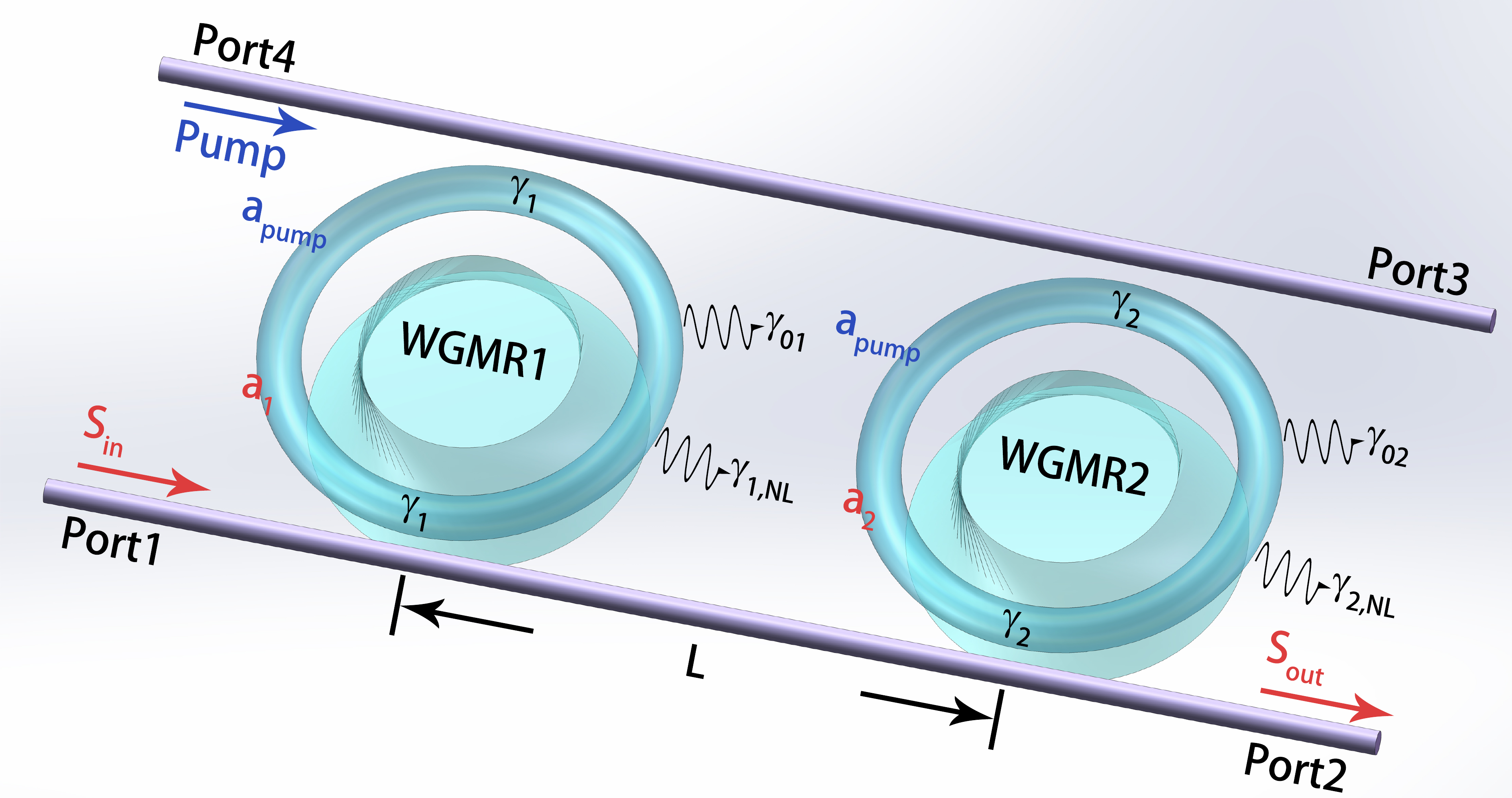}
	\caption{The Schematic diagram of APT symmetric system. WGMR 1 and WGMR 2 are indirectly coupled by two common waveguides, which plays a key role in the implementation of APT symmetry.}
	\label{Fig(1)}
\end{figure}

APT symmetry can be implemented in a dissipative system by forming a dissipative coupling between two detuned modes. The dissipative couplings can be obtained by connecting two separated WGMRs with two common waveguides as shown in Fig.\ref{Fig(1)}. The WGMRs are of the same material and the renonant frequency $\omega_{0j} (j=1,2)$ differs. As $L$ is much larger than the cavities’ radii, the direct coupling of two WGMRs can be neglected. The modal fields in two WGMRs are governed by the coupled-mode equations 
\begin{subequations}
		\label{1}
		\begin{eqnarray}
			\frac{da_{1}}{dt} = &- i\omega_{1}a_{1} - \Gamma_{1}a_{1} - ik\omega_{1}\left| a_{1} \right|^{2}a_{1}  - ga_{2}\label{1a} \\ &-\sqrt{\gamma_{1}}S_{in} \notag\\ 
			\frac{da_{2}}{dt} = &- i\omega_{2}a_{2} - \Gamma_{2}a_{2} - ik\omega_{2}\left| a_{2} \right|^{2}a_{2} - ga_{1}\label{1b}  \\& -\sqrt{\gamma_{2}}e^{i\theta_{d}}S_{in}\notag  
		\end{eqnarray}
	\end{subequations}
where $a_{j}(j=1,2)$ corresponds to the modal field amplitude in each cavity with $|a_j|^{2}$ normalized to the intracavity intensity. $\omega_{j}=\omega_{0j}-\omega$ is the frequency detuning of the j-th WGMR without nonlinear effect, where $\omega$ is the frequency of the input light. $\Gamma_{j}=\gamma_{0j}+2\gamma_{j}+\gamma_{j,NL}-G_{j}$ is the total loss rate of the j-th WGMR, where $\gamma_{0j}$ represents the intrinsic loss rate and  $\gamma_{j}$ corresponds to the coupling loss rate. $\gamma_{j,NL}$ is the loss rate lifted by nonlinear effects. $G_{j}$ is the external gain rate of the j-th WGMR. $k=\frac{2\pi\chi^{(3)}}{n^{2}}$ is the Kerr coefficient. $n$ is the linear refractive index and $\chi^{(3)}$ is the third order nonlinear susceptibility, which are considered as a constant in our scheme. The dissipative coupling strength is $g = e^{i\theta_{d}}\sqrt{\gamma_{1}\gamma_{2}}$. The propagating phase factor $\theta_{d}= \beta_{d}(\omega)L$, where $\beta_{d}(\omega)$ is the waveguide dispersion relationship and $L$ denotes waveguides' length between two WGMRs\cite{xiao2010asymmetric}. $S_{in}$ is the amplitude of input light. Transmission amplitude at Port 2 is defined as $t=\frac{S_{out}}{S_{in}}$ and the transmission rate is $T=|t|^{2}$
	
The eigenvalues of Eq.(\ref{1}) reads
\begin{equation} \label{2}
	\begin{split}
		\omega_{\pm} =& \frac{1}{2} \left( \left( {1 + k\left| a_{1} \right|^{2}} \right)\omega_{1} + \left( {1 + k\left| a_{2} \right|^{2}} \right)\omega_{2}\right) \\&- \frac{i}{2}\left( {\Gamma_{1} + \Gamma_{2}} \right)  \pm\frac{1}{2} \sqrt{\Delta} 
	\end{split}
\end{equation} 
where $\Delta =  \left( \left( 1 + k\left| a_{1} \right|^{2} \right)\omega_{1} - {\left( 1 + k\left| a_{2} \right|^{2} \right)\omega}_{2} + i\left. \left( {\Gamma_{1} - \Gamma_{2}} \right) \right)^{2} \right.\\-4g^{2}$. If $\Delta < 0$, two eigenmodes have the same resonant frequency and different linewidth and the system will be in APT symmetric phase; if $\Delta > 0$, the two eigenmodes have equal linewidth and divergent resonant frequency,  and the system will be in APT symmetry broken phase.
$\Delta=0$ is the location of the EP of the system

\section{Linear APT Symmetry}\label{SEC:3}	To analyze the response of this arrangement under linear conditions, we assume the light intensity is small,i.e.,$|a_{1}|^{2}, |a_{2}|^{2} \sim 0$. Under these assumptions, Kerr effect can be neglected. Hence this regime could be properly described by linearizing Eq.(\ref{1}),e.g.,
	\begin{subequations}
		\label{3}
		\begin{eqnarray}
			\frac{da_{1}}{dt} =& - i\omega_{1}a_{1} - \Gamma_{1}a_{1} - ga_{2} - \sqrt{\gamma_{1}}S_{in}\label{3a} \\ 
			\frac{da_{2}}{dt} =& - i\omega_{2}a_{2} - \Gamma_{2}a_{2} - ga_{1} - \sqrt{\gamma_{2}}e^{i\theta}S_{in} \label{3b} 
		\end{eqnarray}
	\end{subequations}
The eigenvalues of the system in the linear regime read 
\begin{equation}
	\label{4}
	\begin{split}
		\omega_{\pm} =& \frac{1}{2} \left( {\omega_{1} + \omega_{2} - i\left( {\Gamma_{1} + \Gamma_{2}} \right)} \right)\\& \pm \sqrt{ - 4g^{2} + \left( \left( {\omega_{1} - \omega_{2} + i\left. \left( {\Gamma_{1} - \Gamma_{2}} \right) \right)^{2}} \right. \right.} 
	\end{split}
\end{equation}
	
It is clear that $\Gamma_{1}=\Gamma_{2}$ and $\theta_d=\pi$ are necessary to construct APT symmetry in the system\cite{yang2017anti}. Hence, the eigenvalues could be further simplified into 
	\begin{equation}
		\label{5}
		\omega_{\pm} = \omega_{0} - i\Gamma \pm \sqrt{\delta^{2} - g^{2}} 
	\end{equation}
	
Where $\Gamma=\Gamma_{1}=\Gamma_{2}$. $\omega_{0}=\frac{1}{2}(\omega_{1}+\omega_{2})$ and $\delta=\frac{1}{2}(\omega_{1}-\omega_{2})$ define the common- and differential-mode detunings, respectively. In this respect, the two regimes of linear APT symmetry can be identified by whether $g^{2} \gtrless \delta^{2}$. In the first case where $g^{2} > \delta^{2}$, the system remains in APT symmetry unbroken regime. Thus, modal amplitudes in each WGMR are equal and the phase factor $\theta_{l}$ between them is given by $\sin \theta_{l}=-\frac{\delta}{g}$\cite{li2020real,de2019high,yang2017anti}.
	
On the other hand, when $\delta^{2}$ increases beyond $g^{2}$, a spontaneous phase transition would occur and drive the system into APT symmetry broken regime. As opposed to the hallmarks in the unbroken phase, the modal field amplitudes in two WGMRs are unequal, and they are phase shifted by $\theta_{l}=\frac{\pi}{2}$ \cite{li2020real,de2019high,yang2017anti}.  
	
The EP of linear APT symmetric system locates at $\delta^{2}=g^{2}$, which is determined by differential-mode detuning $\delta$ alone (consider $g$ as a constant). As $\delta$ can be easily affected by implementation errors and nonlinear effects, it is difficult to maintain the linear system at its EP precisely. 
	
\begin{figure}[htbp]
		\centering
		\includegraphics[width=8.5cm]{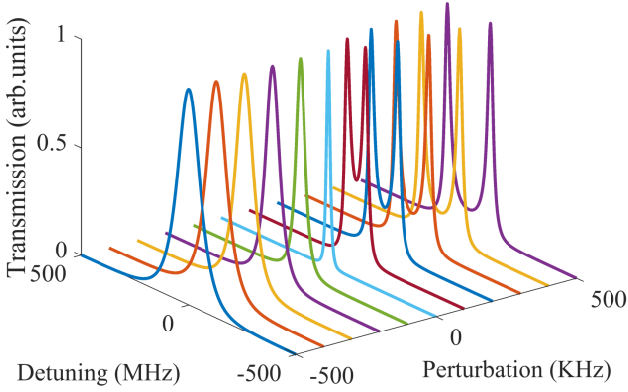} 
		\caption{Transmission Spectrum in the proximity of the linear EP. Perturbations can induce the breakdown of APT symmetry and raise real splitting in transmission spectrum} 
		\label{Fig(2)}
	\end{figure}

Another feature of such EPs is their tremendous response towards external perturbations. Then, we will analyze the response to perturbations of the linear EPs. Assuming the linear system is at its EP initially, and a perturbation $\varepsilon$ influences resonant frequency of WGMRs. The frequency detuning of WGMR1 becomes $\omega_{1}(\varepsilon)=\omega_{1}+m\varepsilon, m\in R$ and that of WGMR2 is $\omega_{2}(\varepsilon)=\omega_{2}+n\varepsilon, n\in R$. Assuming $m-n=2$, $\delta$ changes into $\delta+\varepsilon$. Transmission spectrum in the proximity of the linear EP is depicted in Fig.(\ref{Fig(2)}). In the APT symmetry broken phase, a real splitting appears in the spectrum and the distance between transmission peaks is determined by the scale of perturbations. The parameters of stimulation in Fig.(\ref{Fig(2)}) are $\delta=g=-10 GHz, \Gamma=150 MHz$.
	
	We introduce the frequency splitting in APT symmetry broken regime, defined as $\Delta\omega =\omega_{+}-\omega_{-}$, as a criterion to appraise the responding ability of the system. A larger frequency splitting towards perutrbation indicates better responding ability of the system. The frequency splitting in linear condition is    
	\begin{equation}
		\label{6}
		\Delta\omega_{L}=2\sqrt{(\delta+\varepsilon)^{2}-g^{2}} 
	\end{equation}
	When $\varepsilon\ll 1$, $\Delta\omega_{L}\approx2\sqrt{2|g|\varepsilon}$, which is proportional to $\sqrt{\varepsilon}$. Therefore, the responding ability of such EP system is greatly enhanced compared with that of conventional system, whose response is proportional to $\varepsilon$.

	\section{EP of Nonlinear APT Symmetry} \label{SEC:4}
	As fields in APT-coupled configuration start to grow, Kerr effect comes into play, as described by Eq.(\ref{1}). Yet, Kerr effect’s influence is beyond the common-mode detuning of the system, it might be able to trigger the spontaneous phase transition of APT symmetry, as we will show later. 
	
	Suppose that $a_{2} = \rho e^{i\theta_{k}}a_{1}$, with $\rho > 0,\theta_{k}\in \left[-\frac{\pi}{2},\frac{\pi}{2}\right]$ and solve Eq.(\ref{1}) in the stationary conditon $\frac{d}{dt}a_{1} = \frac{d}{dt}a_{2} = 0$, which reads
	\begin{subequations}
		\label{7}
		\begin{eqnarray}
			&0 = - i{\omega}_{1}a_{1} - \Gamma a_{1} - ik\omega_{1}\left| a_{1} \right|^{2}a_{1} - ga_{2}\label{7a} \\ 
			&0 = - i{\omega}_{2}a_{2} - \Gamma a_{2} - ik\omega_{2}\left| a_{2} \right|^{2}a_{2} - ga_{1} \label{7b} 
		\end{eqnarray}
	\end{subequations}
	After dividing Eq.(\ref{7a}) by $a_{1}$ and Eq(\ref{7b}) by $a_{2}$, upon subtraction comes that
	\begin{widetext}
		\begin{equation}
			\label{8}
			\begin{split}
				g\left( {\rho - \frac{1}{\rho}} \right)cos\theta_{k} + i\left( {2\delta + k\left( {\omega_{1}\left| a_{1} \right|^{2} - \omega_{2}\left| a_{2} \right|^{2}} \right) + g\left( {\rho + \frac{1}{\rho}} \right)sin\theta_{k}} \right) = 0 
			\end{split}
		\end{equation}
	\end{widetext}
	
	Eq.(\ref{8}) can be solved for the real part and imaginary parts. To vanish the real part of Eq.(\ref{8}), $\rho-\frac{1}{\rho}=0$ or $\cos \theta_{k}=0$ must be satisfied. If $\rho-\frac{1}{\rho}=0$, the system is in APT symmetric regime, while $\cos \theta_{k}=0$ corresponds to the nonlinear system in APT symmetry broken regime, and the EP locates at where the two conditions are satisfied simutaneously, as we will prove later.
	\subsection{Nonlinear APT Symmetric Regime} 
	
	If $\rho-\frac{1}{\rho}=0$, the intensity in two WGMRs is identical to each other. Then, we can assume $|a_{1}|^{2}=|a_{2}|^{2}=|a|^{2}$. By solving the imaginary part of Eq.(\ref{8}) based on this assumption, the relationship between light intensity $|a|^{2}$ and phase factor $\theta_{k}$ can be obtained as 
	\begin{equation}
		\label{9}
		\left| a \right|^{2} = - \frac{1}{k}\left( \frac{g}{\delta}sin\theta_{k} + 1 \right) 
	\end{equation}  
	Owing to the fact that light intensity should be positive in physics perspective and $\theta_{k}\in R$, $\left| a \right|^{2} \leq -\frac{1}{k}\left( {1+\frac{g}{\delta} } \right)$ (given that $k,\delta,g<0$). Under these conditions, $\Delta=4g^{2}(\sin^{2}\theta_{k}-1)\leq0$ ,suggesting that the system is in APT symmetric regime. The threshold of intracavity intensity lies at $|a_{th}|^2=- \frac{1}{k}\left( 1+\frac{g}{\delta} \right)$. At this point, the nonlinear phase factor reaches its maximum at $\sin \theta_{k}=1$. If the intracavity intensity continues to grow, $\theta_{k}$ will no longer be real and APT symmetry will break down.
	
	Besides, when the intracavity intensity is weak and Kerr effect is negligible, ${\sin\theta_{k}} \approx  -\frac{\delta}{g}$ and $\omega_{\pm} \approx \omega_{0} -i \Gamma \pm \sqrt{\delta^{2}-g^{2}  } $, the system will degenerate into the linear case.
	\subsection{Nonlinear APT Symmetry Broken Regime} 
	When  $\cos \theta_{k}=0$, $a_{1}$ and $a_{2}$ are out of phase by $\frac{\pi}{2}$. Then Eq(\ref{10}) is obtained through solving the imaginary part of Eq.(\ref{8})
	\begin{equation}
		\label{10}
		\left| a_{1} \right|^{2} = \frac{g\left( {1 + \rho^{2}} \right) + 2\delta\rho}{-k\rho\left( \omega_{1} - \rho^{2}\omega_{2} \right)}
	\end{equation}
	By taking Eq(\ref{10}) into  Eq(\ref{2}), we find that $\Delta=4g^{2}\left( {\rho - \frac{1}{\rho}} \right)^{2}\geq 0$.
	\begin{figure}[htbp]
		\centering
		\includegraphics[width=8.5cm]{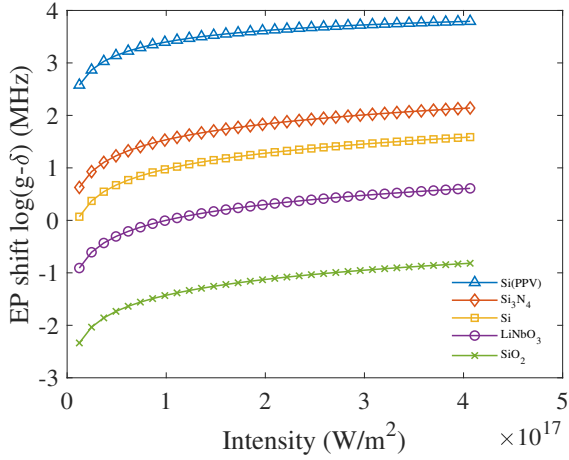} 
		\caption{EP shift versus light intensity. Coupling coefficient $g$ is taken as $-10 GHz$. EP shift is defined as $g-\delta$. Materials with larger Kerr coefficient possess larger Kerr frequency shift, which could move the EP with less intensity.The Kerr coefficient utilized in the simulation is $k_{Si(PPV)}=2\times10^{-15}m^{2}/W,k_{Si_{3}N_{4}}=2.17\times10^{-17}m^{2}/W,k_{Si}=6\times10^{-18}m^{2}/W,k_{LiNbO_{3}}=6.31\times10^{-19}m^{2}/W$ and $k_{SiO_{2}}=2.36\times10^{-20}m^{2}/W$\cite{rukhlenko2009analytical,lin2015si,zhu2019nonlinear,wang2019monolithic,razdolskiy2011hybrid}} 
		\label{Fig(3)}
	\end{figure} 
	As $\Delta$ is non-negative with $\cos \theta_{k}=0$, the system is in APT symmetry broken phase. Threshold of this regime lies in $\rho=1$, where $\Delta$ equals to zero.

	The EP of nonlinear APT symmetry locates at  
	\begin{equation}
		\label{11}
		{~\left( {1 + k\left| a \right|^{2}} \right)}^{2}\delta^{2} - g^{2} = 0 
	\end{equation}
	where both $\rho-\frac{1}{\rho}=0$ and $\cos \theta_{k}=0$ are satisfied. It can be clearly seen in Eq.(\ref{11}) that EPs of the nonlinear system are correlated to both differential-mode detuning $\delta$ and intracavity light intensity $|a|^2$. Via Kerr effect, light intensity becomes an additional degree of freedom to control the phase transition in the system. The relationship between EP shift $g-\delta$ and intracavity intensity $|a|^{2}$ is illustrated in Fig(\ref{Fig(3)}). The EP shift $g-\delta$ increases with the increment of intracavity intensity $|a|^{2}$. This dependence on intensity can be utilized to manipulate the position of EP to overcome imperfections resulting in the mismatch of $\delta$ and $g$ and locate the system at its EP precisely. 
	
	In addition, Kerr coefficient determines the scale of Kerr shift when intracavity intensity is limited. As seen in Fig(\ref{Fig(3)}), materials with larger Kerr coefficient can have larger EP shift and compensate the mismatch of $\delta$ and $g$ with less power. Hence, materials with larger Kerr coefficient are preferable for this nonlinear manipulation. It is also worth noting that the direction of Kerr shift is often red, $\delta$ should be blue-detuned from $g$ initially to ensure the effectiveness of the scheme.

	\section{Frequency Splitting of Nonlinear APT symmetry}\label{SEC:5}
	Nonlinear effect could not only shift EPs of APT symmetry, but also affect the responding ability of the system. In this section, analyses on how nonlinear effct influence the transmission spectrum and frequency splitting have been made.
	\subsection{Influence of SPM}
\begin{figure}[htbp]
		\centering	\includegraphics[width=8.5cm]{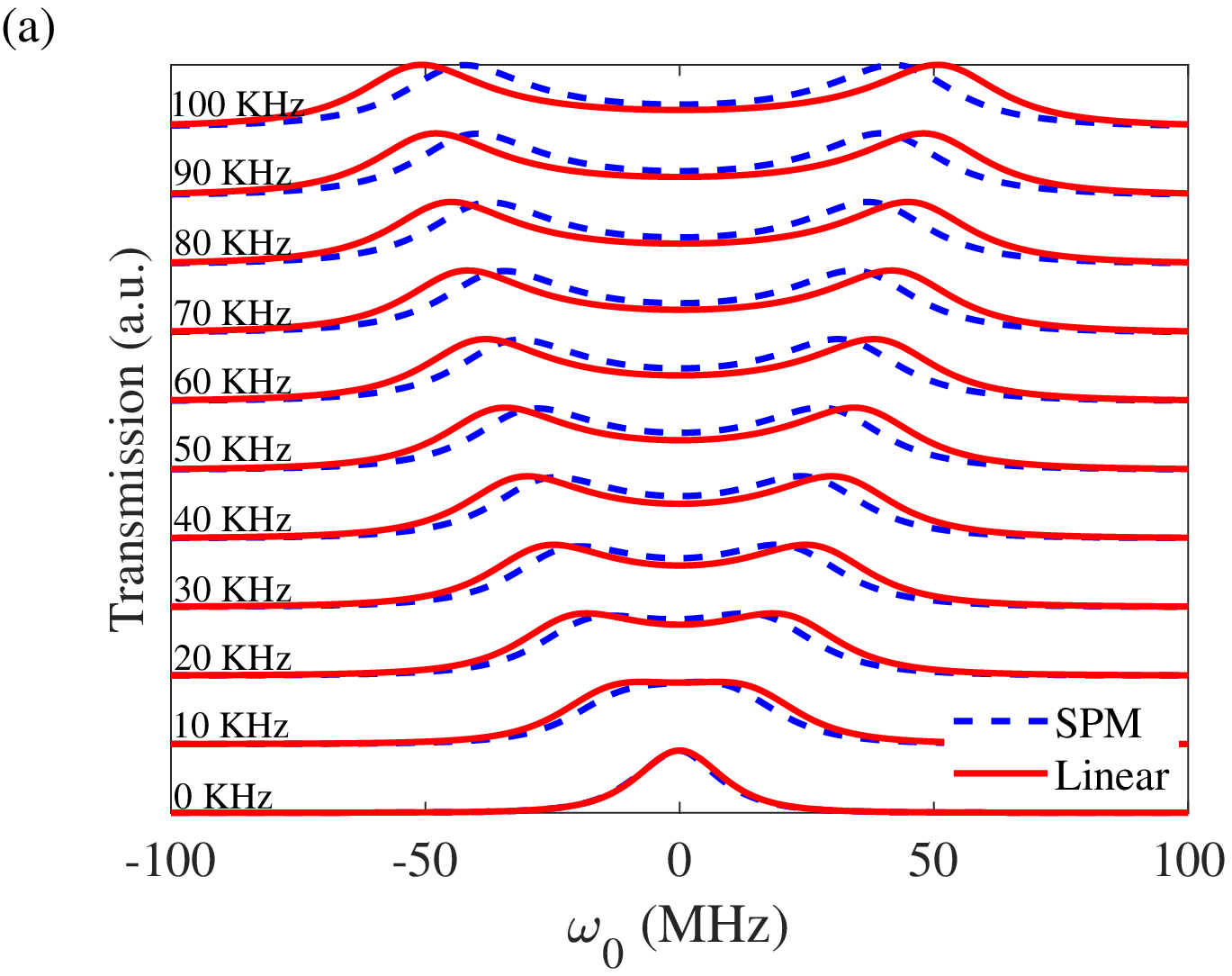}
		\centering	\includegraphics[width=8.5cm]{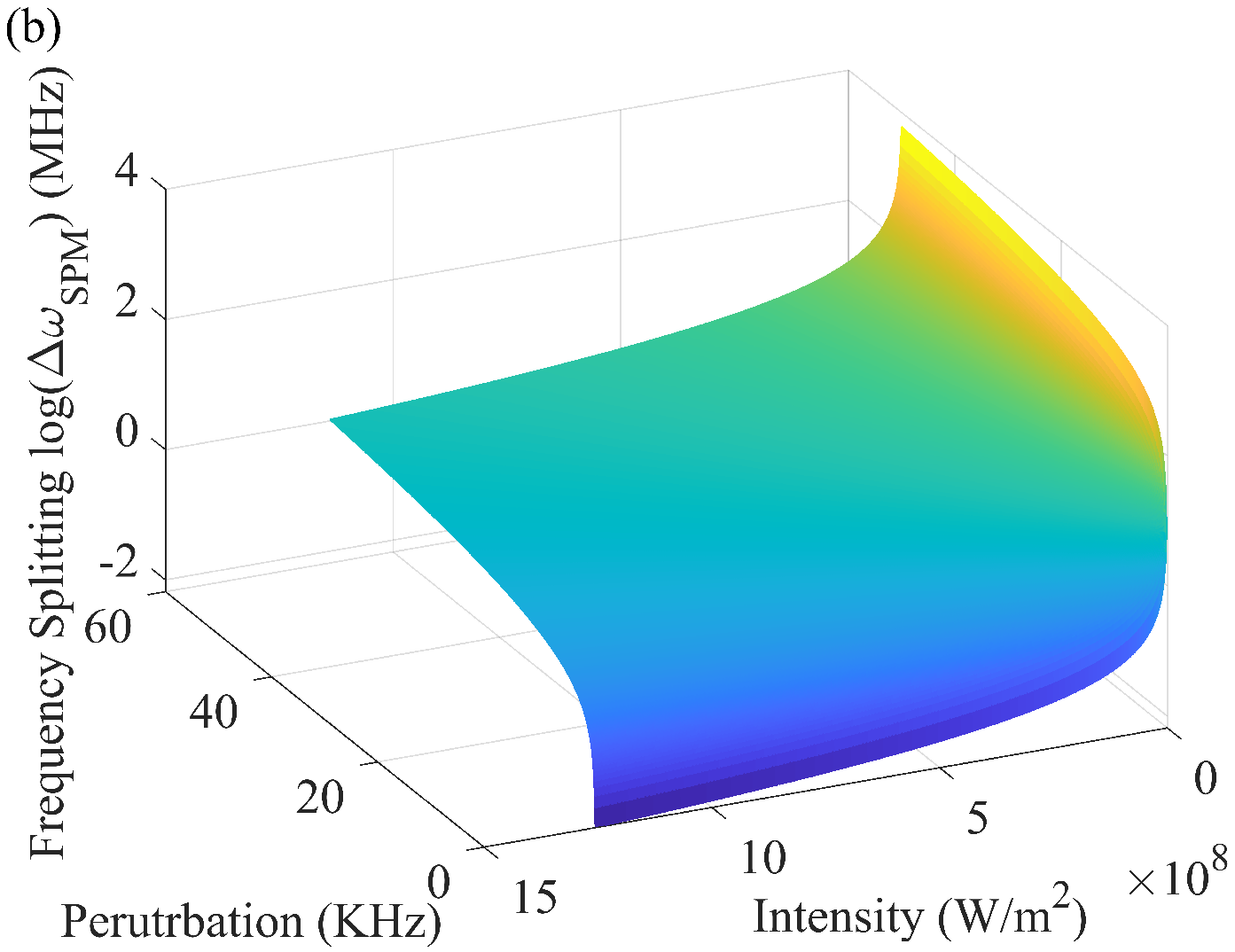}
		\caption{(a) Transmission spectrum of SPM and linear condition. Frequency splitting of SPM (blue dashed curve) is much weaker than that in linear condition(red solid curve).The input intensity of probe light $|S_{in}|^{2}=7\times10^{6} W/m^{2}$. The total loss rate is $\Gamma=150 MHz$ and the Kerr coefficient is $k=6\times10^{-18}m^{2}/W$. The intracavity strength $|a|^2$ and $\delta$ could be caculated from Eq.(\ref{7}).  (b) Frequency splitting under SPM versus perturbation and light intensity. Stronger perturbation and weaker intensity will lead to larger frequency splitting.}
		\label{Fig(4)}
	\end{figure}

	In this part, probe light is strong enough to stimulate Kerr effect itself (assuming the light intensity is not strong enough to trigger bistability of optical resonant mode\cite{grudinin2009thermal,carmon2004dynamical,shim2016nonlinear}). Transmission spectra versus different perturbations could be obtained by numerically solving Eq(\ref{7}), and are shown in Fig(\ref{Fig(4)}a). 
	
	In Fig.(\ref{Fig(4)}a), it can be clearly seen that the frequency splitting gets larger with an increase of perturbation in both conditions. However, a tiny difference exists between the resonant frequency of SPM and linear case when there is no perturbation, since Kerr effect causes red shift of resonant frequency. Moreover, the SPM scheme (blue dashed curve) exhibits smaller frequency splitting than that of linear case (red solid curve). We can gain insight into the reason for the decrease through the frequency splitting under SPM, which reads
	\begin{widetext}
	\begin{equation}\label{12}
		\Delta\omega_{SPM}=2\sqrt{\left((\delta+\varepsilon)(1+k|a_1|^{2})-k\omega_{2}(\varepsilon)(|a_{2}|^{2}-|a_{1}|^{2})\right)^{2}- g^{2}}
	\end{equation}
	\end{widetext}
	When the system is in APT symmetry broken phase, intracavity light intensity is unequal and an excess frequency shift $\omega_{a} =k\omega_{2}(\varepsilon)(|a_{2}|^{2}-|a_{1}|^{2})$ will be lifted by the asymmetric light intensity. The direction of $\omega_{a}$ is against $\varepsilon $, which means it tends to pull the system back to its EP. Therefore, $\Delta\omega_{SPM}$ is smaller than $ \Delta\omega_{L}$ and the responding ability of the system has been weakened by SPM.
	
	Fig(\ref{Fig(4)}b) depicts frequency splitting under SPM versus perturbations and  intracavity intensity. It can be observed that frequency splitting increases when the perturbation gets larger. However, stronger intracavity intensity will bring about larger frequency shift $\omega_{a}$ and become a detriment to the responding ability of the system, which is unwanted in detecting tiny perturbations. Hence, we put forward the scheme of XPM in order to solve the decrease of frequency splitting.
\subsection{Influence of XPM}
\begin{figure}[htbp]
	\centering
	\includegraphics[width=8.5cm]{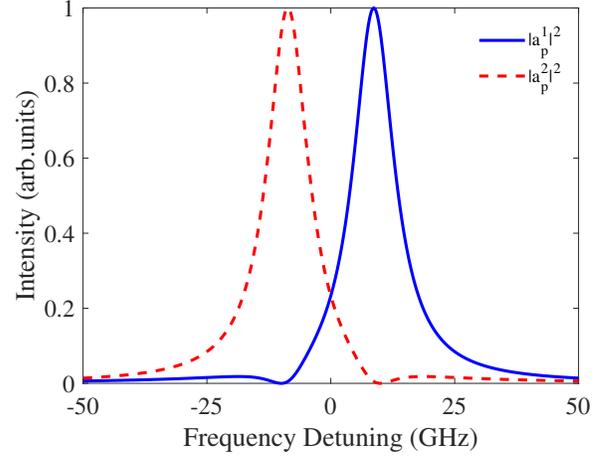} 
	\caption{Intensity of pump light in WGMRs in different detunings. As the frequency of pump light varies, intensity of pump light varies in each WGMR. The resonant frequency is chosen as the frequency where intracavity pump light is the same. the loss rate of the pump light is $\Gamma_{p}=10 GHz$. } 
	\label{Fig(5)}
\end{figure}

In this part, an extra pump light is introduced into the system to induce high intensity. Frequency of the pump light is far-detuned from the probe light, thus it cannot be observed in transmission spectrum directly. Importantly, the intensity of the probe light is too weak to induce any significant Kerr shift on its own\cite{PhysRevLett.124.223901}. Hence, coupled-mode euqations under XPM read 
	\begin{subequations}
		\label{13}
		\begin{eqnarray}
			\frac{da_{1}}{dt}=&-i\omega_{1}a_{1} - \Gamma a_{1} - 2ik\omega_{1}\left| a^{1}_{p} \right|^{2}a_{1} - ga_{2}\label{13a}\\& - \sqrt{\gamma_{1}}S_{in}\notag  \\
			\frac{da_{2}}{dt}=&-i\omega_{2}a_{2} - \Gamma a_{2} - 2ik\omega_{2}\left| a^{2}_{p} \right|^{2}a_{2} - ga_{1} \label{13b}\\&- \sqrt{\gamma_{2}}e^{i\theta_{d}}S_{in} \notag
		\end{eqnarray}
	\end{subequations}
The factor of 2 arises due to XPM in the WGMRs. $|a^{1}_{p}|^{2}$ and $|a^{2}_{p}|^{2}$ are intensity of pump light in the WGMRs, respectively. Fig.(\ref{Fig(5)}) demonstrates the reliance of the pump intensity inside each WGMR on the pump frequency. Therefore, intracavity pump intensity can be modulated by both input pump power and pump frequency. 

	Note that the eigenvalues of system under XPM read
	\begin{equation}
		\omega_{\pm}^{'} = \frac{1}{2}\left\lbrack \left( {\left( {1 + 2k\left| a_{p}^{1} \right|^{2}} \right)\omega_{1} + {\left( {1 + 2k\left| a_{p}^{2} \right|^{2}} \right)\omega}_{2} - 2i\Gamma} \right) \pm \sqrt{\Delta^{'}} \right\rbrack
		\label{14}
	\end{equation}
	where $\Delta^{'}~ = - 4g^{2} + [ 2( 1 + 2k\left| a_{p}^{1} \right|^{2} )\delta +2k( \left| a_{p}^{1} \right|^{2} - \left| a_{p}^{2} \right|^{2} )\omega_{2}]^{2}$. The EP of the system is situated at $\Delta^{'}=0$. In the SPM scheme, intracavity intensity must be identical at its EP. However, XPM scheme allows the APT symmetric system to arrive at its EP with nonidentical intracavity intensity. In this scheme, both the input power and the frequency of the pump light can be applied to modulate the EP. The difference of pump intensity in two WGMRs $ \left| a_{p}^{1} \right|^{2} - \left| a_{p}^{2} \right|^{2}$, determined by pump frequency, can influence the position of the EP as well as the intensity of pump light $ \left| a_{p}^{1} \right|^{2}$. Besides, as increment of pump light would not trigger bistability of resonant mode, the restraint on light intensity in order to prevent bistability is removed and Kerr shift induced by pump light can be much larger than that under SPM.  
	
	Then, we will analyze the responding ability of EPs under XPM. Provided that intensity of pump light remains as a constant during the perturbation, frequency splitting under XPM is
 \begin{widetext}
 	\begin{equation}
		\label{15}
		\Delta\omega_{XPM} = \sqrt{- 4g^{2} + [ ( 1 + 2k\left| a_{p}^{1} \right|^{2} )( \omega_{1} + m\varepsilon ) -( 1 + 2k\left| a_{p}^{2} \right|^{2} )( \omega_{2} + n\varepsilon)]^{2}}
	\end{equation} 
\end{widetext}

	Note that contrasting to linear condition and SPM, frequency splitting under XPM $\Delta\omega_{XPM}$ relies on $m,n$, which is determined by the perturbation itself. In the former case, the coefficient of $m, n$, which is determined by Kerr nonlinearity, is almost identical. Thus, they appear in the form of $m-n$ when considering the frequency splitting. With the assumption of $m-n=2$, they can hardly have separate influence on the frequency splitting albeit WGMRs have distinct response towards perturbations. However, the coefficient of $m,n$ can have significant difference under XPM due to the diffrent intracavity pump intensity. How WGMRs respond to perturbations is now essential to the frequency splitting under XPM. As the Kerr coefficient $k$ is often negative and frequency shift of resonant mode is often red, suppressing Kerr nonlinearity inside the WGMR with more acute response will be helpful to enhance the frequency splitting. 
	
\begin{figure}[htbp]
		\centering
		\includegraphics[width=8.5cm]{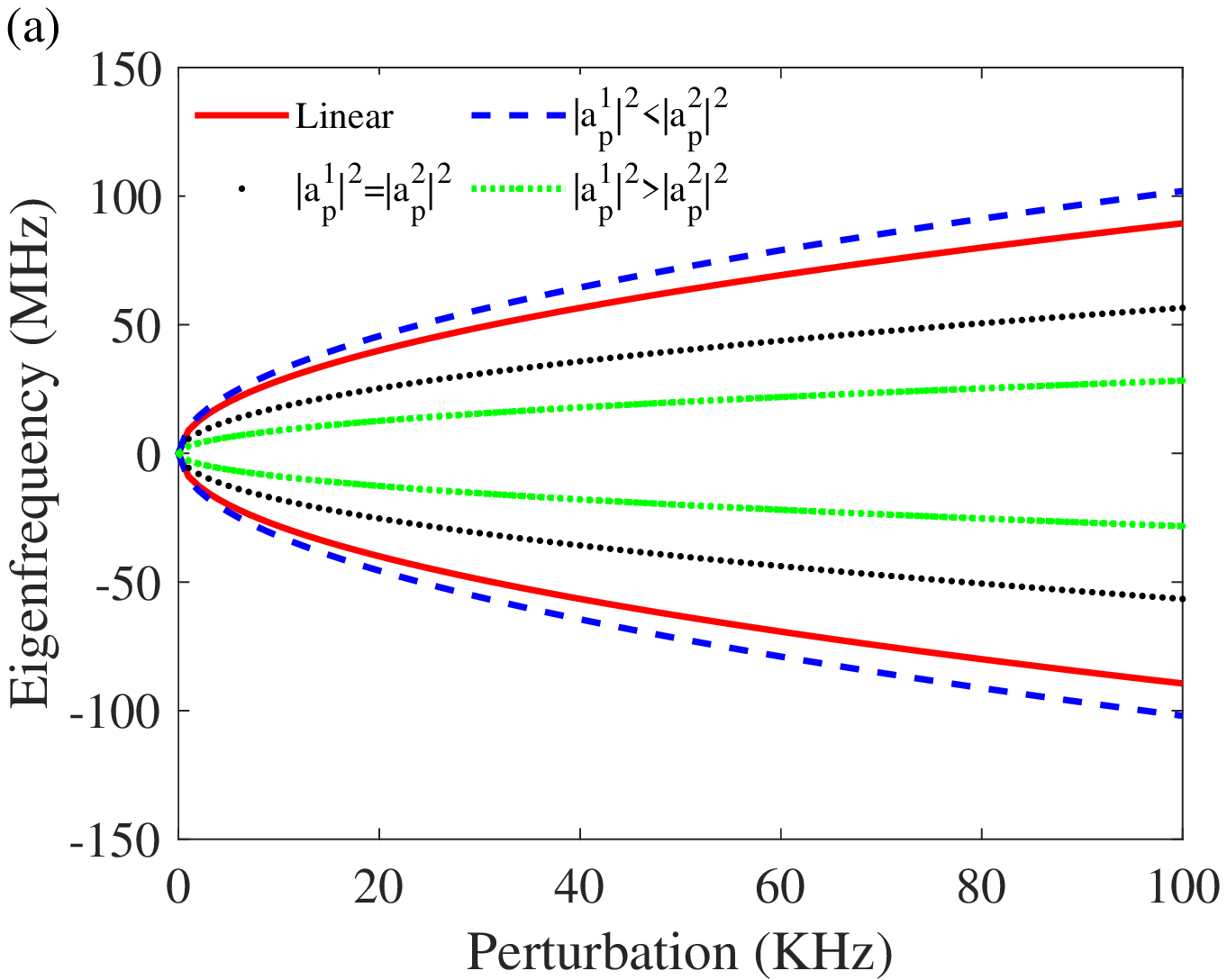} 
		\includegraphics[width=8.5cm]{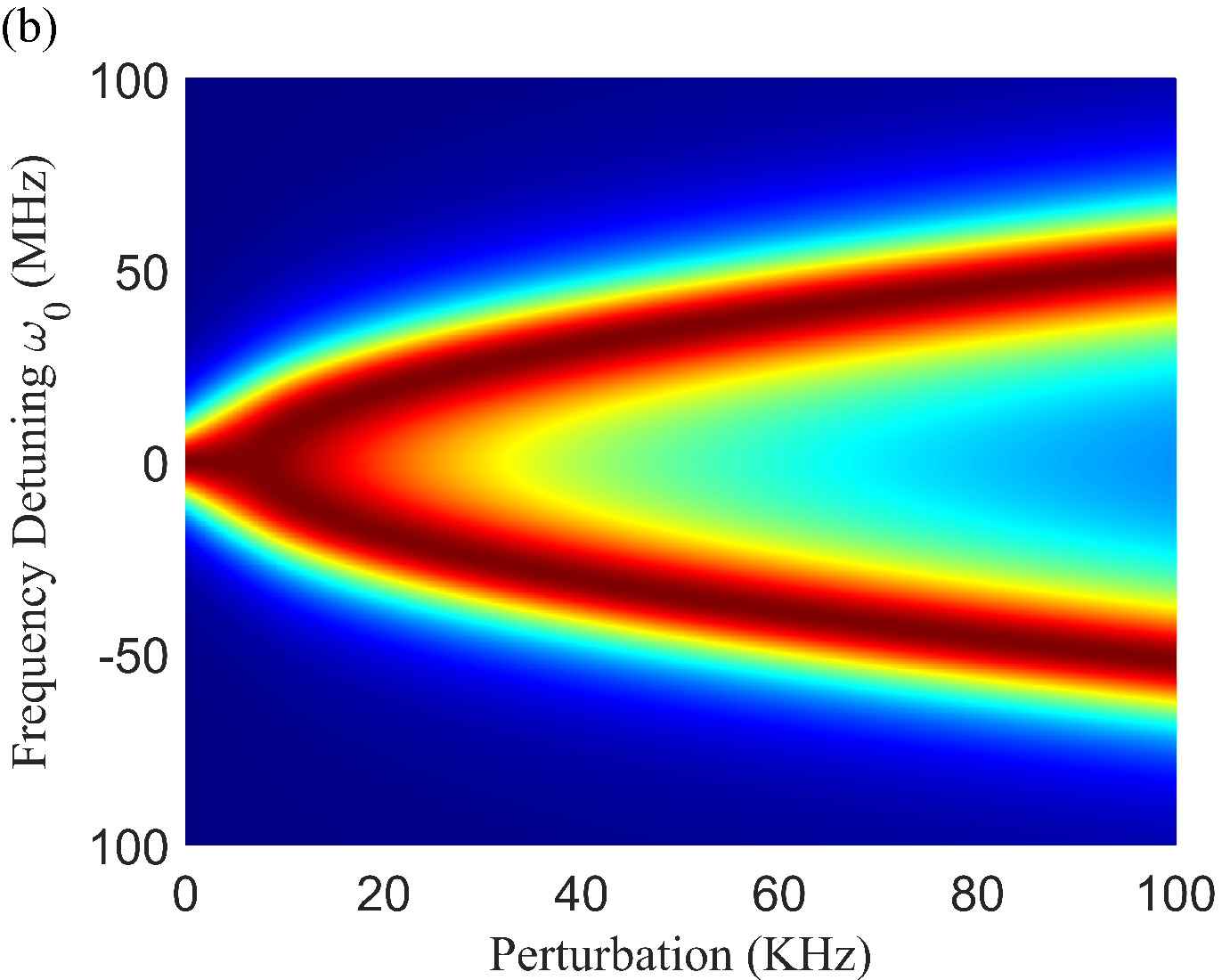} 
		\caption{(a) Eigenfrequency splitting under different pump condition. Eigenfrequency splitting depends on the pump condition and $\Delta\omega_{XPM}$ (blue dashed curve) can be larger than $\Delta\omega_{L}$ (red solid curve) by judiciously choosing the pump intensity. (b) Trajectory of the transmission spectrum under XPM as a function of perturbations. The splitting of transmission peak is consistent with the splitting of eigenfrequency.  Parameters taken in the simulation is $m=3,n=1$,$|a_{p}^{1}|=0,|a_{p}^{2}|=5\times10^{16}W/m^{2}$ for $|a^{1}_{p}|^{2}<|a^{2}_{p}|^{2}$,$|a_{p}^{1}|=|a_{p}^{2}|=5\times10^{16}W/m^{2}$ for $|a^{1}_{p}|^{2}=|a^{2}_{p}|^{2}$,$|a_{p}^{2}|=5\times10^{16}W/m^{2},|a_{p}^{2}|=0$ for $|a^{1}_{p}|^{2}>|a^{2}_{p}|^{2}$. } 
		\label{Fig(6)}
	\end{figure}

	In Fig.(\ref{Fig(6)}a) we present a series of eigenfrequency splitting versus perturbations. It clearly exhibits that frequency splitting under XPM depends on the relative strength of pump intensity in each WGMR. When Kerr nonlinearity in WGMR with more drastic response is repressed (the blue dashed curve, assuming $|m|>|n|$), enhancement of frequency splitting can be expected. The increase of frequency splitting reaches its maximum at $\left| a_{p}^{1} \right|^{2}=0$, when only WGMR 2 is pumped. Fig.(\ref{Fig(6)}b) shows the transmission peak trajectory of the probe ligth, which is consistent with the paths of eigenfrequencies expressed by Eq.(\ref{14}). This confirms the enhanced requency splitting can be observed in the spectra directly. 
	
	\section{Thermal effect In The System}\label{SEC:6}
	High pump intensity increases light absorption of the cavity, causing significant thermal effect as well as Kerr effect. In this section, we will investigate the influence of thermal effect induced by high intensity and other mechanisms causing the variance of temperature will be neglected.
	
	When taking thermal effect into consideration, the dynamic evolution of the system under XPM can be described by 
	\begin{subequations}
		\label{16}
		\begin{eqnarray}
			\frac{da_{1}}{dt} &= - i\omega_{1}a_{1} - \Gamma_{1}a_{1} - ik\omega_{1}\left| a_{p}^{1} \right|^{2}a_{1}- \omega_{1}\beta\theta_{1}a_{1}\label{16a} \\ & - ga_{2}-\sqrt{\gamma_{1}}S_{in}\notag \\ 
			\frac{da_{2}}{dt} &= - i\omega_{2}a_{2} - \Gamma_{2}a_{2} - ik\omega_{2}\left| a_{p}^{2} \right|^{2}a_{2} - \omega_{2}\beta\theta_{2}a_{2} \label{16b} \\& - ga_{1}-\sqrt{\gamma_{2}}e^{i\theta_{d}}S_{in} \notag \\
			&\frac{d\theta_{1}}{dt} = \frac{\delta_{\theta}\nu}{\beta}\left| a_{p}^{1} \right|^{2} - \delta_{\theta}\theta_{1} \label{16c}\\
			&\frac{d\theta_{2}}{dt} = \frac{\delta_{\theta}\nu}{\beta}\left| a_{p}^{2} \right|^{2} - \delta_{\theta}\theta_{2} \label{16d}
		\end{eqnarray}
	\end{subequations}  
	$\beta = \frac{1}{n}\frac{dn}{dT} + \alpha$ is a temperature coefficient which contains both thermal index change and thermal expansion. $\theta_{j} (j=1,2)$ is the temperature difference between the mode volume of the j-th WGMR and the ambient. $\delta_{\theta}$ is determined by thermal diffusion coefficient and half thickness of the mode in radial direction\cite{grudinin2009thermal}. $\nu=\frac{\beta\alpha_{A} nc}{4\pi C_{h}\rho\delta_{\theta}}$ defines the frequency shift coefficient due to thermal effect, where $c$ is light speed in vacuum, $\alpha_{A}$ is the absorption coefficient of the WGMR, $C_{h}$ is the heat capacity of the WGMR and $\rho$ is the density of the WGMR. 
	\begin{figure}[htbp]
		\centering
		\includegraphics[width=8.5cm]{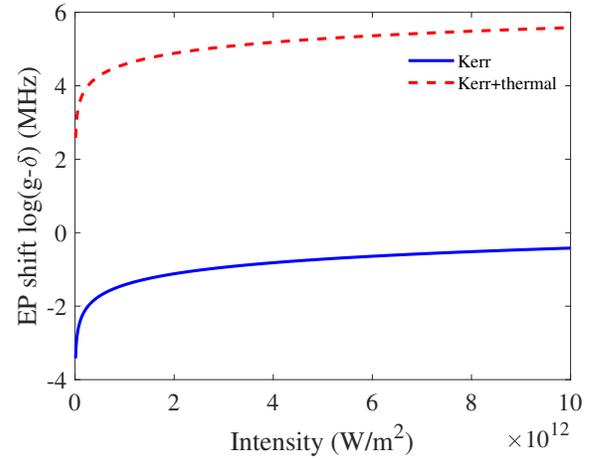} 
		\caption{EP shift versus light intensity. Kerr effect is weak for silica, yet if thermal effect induced by high intensity is considered, EP shift under the same pump intensity can be greatly enhanced.} 
		\label{Fig(7)}
	\end{figure}

	Set all time derivatives in Eq.(\ref{16}) to zero and its static solution comes that 
	\begin{equation}
		\label{17}
		\theta_{j} = \frac{\nu}{\beta}\left| a_{p}^{j} \right|^{2}    (j=1,2)
	\end{equation}
	
	As $\theta_{j}$ is proportional to pump light intensity $|a_{p}^{j}|^{2}$, the frequency shift induced by thermal effect is also proportional to the intensity of pump light in each cavity. The total frequency shift lifted by nonlinear effect in the j-th WGMR is 
	\begin{equation}
		\label{18}
		\Delta\omega_{j,NL}= \mu\omega_{j}|a_{p}^{j}|^2
	\end{equation}
	where $\mu=k+\nu$ is the total nonlinear frequency shift coefficient. As the frequency shift caused by thermal effect is usually in the same direction with Kerr effect, thermal effect raised by high pump intensity could be incorporated with Kerr nonlinearity to manipulate the EP more effectively.
	
	Silica is taken as an example here to demonstrate the influence of thermal effect, as seen in Fig.(\ref{Fig(7)}). $k$ for silica is only $2.36\times10^{-20}m^{2}/W$\cite{zhu2019nonlinear}, while $\nu$ for silica could be up to $ 2.37\times10^{-14}m^{2}/W$\cite{grudinin2009thermal}, which is $6$ orders larger than its Kerr coefficient. In Fig.(\ref{Fig(7)}), EP shift is almost $6$ orders larger with thermal effect. The EP shift $g-\delta$ is greatly enhanced with the help of thermal effect and  manipulation of EPs could be accomplished with less pump intensity.

\section{Conclusion} \label{SEC:7}
We have presented a scheme to modulate EPs in APT symmetry configuration by Kerr effect. The position of EP is manipulated by light intensity, which can also trigger the spontaneous phase transition of APT symmetry. Modulating EPs by SPM will hamper the responding ability of the APT symmetric system. The XPM scheme is then introduced to realize the enhancement of the system's response to detectable perturbations. Moreover, thermal effect induced by high pump intensity can reduce the pump intensity required to manipulate the EP. Our research would be helpful to experimental realization of such APT structure and pave the way for appications of APT symmetry in integrated photonics devices.
	
	\begin{acknowledgments}
		We acknowledge financial support from the National Natural Science Foundation of China (Grants No.61975005, No.11804017, and No.51872010) and Beijing Academy of Quantum Information Sciences (Grant No.Y18G28).
	\end{acknowledgments}
	
	\section*{AUTHOR DECLARATIONS}
	\subsection*{Conflict of Interest}
	The authors have no conflicts to disclose.
	
	\section*{DATA AVAILABILITY}
	
	The data that support the findings of this study are available
	from the corresponding author upon reasonable request.
	\nocite{*}
\providecommand{\noopsort}[1]{}\providecommand{\singleletter}[1]{#1}%
\providecommand{\noopsort}[1]{}\providecommand{\singleletter}[1]{#1}%

\end{document}